\documentclass[times]{nmeauth}

\usepackage{moreverb}

\usepackage[dvips,colorlinks,bookmarksopen,bookmarksnumbered,citecolor=red,urlcolor=red]{hyperref}

\newcommand\BibTeX{{\rmfamily B\kern-.05em \textsc{i\kern-.025em b}\kern-.08em
T\kern-.1667em\lower.7ex\hbox{E}\kern-.125emX}}

\def\volumeyear{2022}

\usepackage{kurz}
\usepackage{amssymb}
\usepackage{mathrsfs}
\usepackage{multirow}    % BY THOMAS
\usepackage{amsmath}
\usepackage{caption}  
\usepackage{stmaryrd}
\usepackage{rotating}
\usepackage{setspace}
\usepackage[numbers,sort&compress]{natbib}

\usepackage{alphalph}
\makeatletter
\newalphalph{\fnsymbolwrap}[wrap]{\@fnsymbol}{}
\makeatother

\begin{document}
%\begin{spacing}{2.5} 

%% Title, authors and addresses

%% use the tnoteref command within \title for footnotes;
%% use the tnotetext command for the associated footnote;
%% use the fnref command within \author or \address for footnotes;
%% use the fntext command for the associated footnote;
%% use the corref command within \author for corresponding author footnotes;
%% use the cortext command for the associated footnote;
%% use the ead command for the email address,
%% and the form \ead[url] for the home page:
%%
%% \title{Title\tnoteref{label1}}
%% \tnotetext[label1]{}
%% \author{Name\corref{cor1}\fnref{label2}}
%% \ead{email address}
%% \ead[url]{home page}
%% \fntext[label2]{}
%% \cortext[cor1]{}
%% \address{Address\fnref{label3}}
%% \fntext[label3]{}

\title{Virtual Displacement based Discontinuity Layout Optimization}

\author{Yiming Zhang\affil{a}\corrauth, Xueya Wang\affil{a}, Xinquan Wang\affil{b,c}, Herbert Mang\affil{b,d,e}\corrauth}

%% use optional labels to link authors explicitly to addresses:
%% \author[label1,label2]{<author name>}
%% \address[label1]{<address>}
%% \address[label2]{<address>}

\address{\affilnum{a}School of Civil and Transportation Engineering, Hebei University of Technology, Xiping Road 5340, 300401~Tianjin,~China \break 
	\affilnum{b}Institute for Mechanics of Materials and Structures (IMWS), Vienna University of Technology,Karlsplatz 13/202, 1040~Vienna, Austria \break
	\affilnum{c}Key Laboratory for Mechanics in Fluid Solid Coupling Systems, Institute of Mechanics, Chinese Academy of Sciences,100190~Beijing,~China \break 
		\affilnum{d}State Key Laboratory for Disaster Reduction in Civil Engineering, Tongji University, Siping Road 1239, 200092~Shanghai,~China \break
		\affilnum{e}Department of Geotechnical Engineering, Tongji University, Siping Road 1239, 200092~Shanghai,~China 
}

\corraddr{Herbert A. Mang and Yiming Zhang\\
	\mbox{~~~~~~~~~~~~~~~~~~~~~~~Email:} Herbert.Mang@tuwien.ac.at, Yiming.Zhang@hebut.edu.cn}

\begin{abstract}
\Large
Discontinuity layout optimization (DLO) is a relatively new upper bound limit analysis method.  Compared to classic topology optimization methods, aimed at obtaining the optimum design of a structure by considering its self-weight, building cost or bearing capacity, DLO optimizes the failure pattern of the structure under specific loading conditions and constraints by minimizing the dissipation energy.  In this work, we present a modified DLO algorithm that contains all of the advantages of DLO.  It is referred to virtual displacement-based discontinuity layout optimization (VDLO).  VDLO takes the stress state of a loaded structure as a snapshot and correspondingly provides the optimum failure pattern, which greatly extends the application potential of DLO.  Numerical examples indicate the effectiveness and flexibility of VDLO.  It is regarded as a highly promising supplemental tool for other numerical methods in element-/node-based frameworks.

\end{abstract}


\maketitle

\vspace{-6pt}

\Large
%Editor: We have provided a PDF that shows the tracked changes in your file as in a Word document. This method makes it easier for you to match the edited file with your original file and make any necessary edits to your file in your LaTeX program. Please let us know if you require further assistance.

\section{Introduction}
\label{sec:it}
Numerical topology optimization methods are broadly utilized to optimize the shapes of structures, considering specific target functions such as: minimization of the self-weight or maximization of the bearing capacity of a structure \cite{Bendsoe:01,linsen2016}.  As a specific type of topology optimization algorithms, layout optimization methods optimize combinations of truss/beam members.  Fortunately, this approach commonly leads to convex optimization problems.  Given the development of modern optimization solvers, layout optimization problems with millions of unknowns can be solved efficiently \cite{Gilbert:04,Gilbert:07}.

The discontinuity layout optimization (DLO) method, first presented in \cite{Smith:01}, optimizes the layout of discontinuities in a loaded structure.  During optimization, DLO introduces millions of potential discontinuities that intersect with each other; the algorithm then seeks to minimize the dissipation energy of the system by optimizing the layout of these discontinuities.  Since its inception, DLO has been successfully used to compute upper bound limits in several geotechnical \cite{Smith:01,Smith:02,Smith:03,Clarke:01,Yiming:17,sunzizheng2019b} and structural \cite{Gilbert:01,Gilbert:03,HeLinWei:01,HeLinWei:02} problems, demonstrating the high efficiency and reliability of this technique for 2D as well as 3D problems \cite{Hawksbee:01,Yiming:12}.  Nevertheless, DLO has some drawbacks. i) It assumes rigid body displacements of the subdomains.  Hence, the elastic properties of the material are not taken into account.  ii) When considering body forces such as gravity, DLO accumulates the total forces over the discontinuities, which is inconvenient.  iii) Despite being a powerful numerical tool for calculating upper bound limit load and potential failure patterns, DLO is not compatible with the pre- and postprocessing steps of some other popular numerical tools incorporated in the finite element method (FEM) framework.

In this work, we present a modified version of the DLO method, named virtual displacement-based discontinuity layout optimization (VDLO), which partly solves the aforementioned problems.  VDLO takes the stress field of the loaded structure as a snapshot and then obtains the upper bound limit load and failure pattern.  Similar to DLO, VDLO also shows weak discretization sensitivity and high efficiency.  Moreover, unlike DLO, VDLO has the following advantages:
\begin{itemize}
	\item
	VDLO takes a stressed body as a snapshot and adopts the stress state/field as input, with the stress field obtained by other numerical tools, considering predefined constitutive relations.  The material can be elastic, elastoplastic, or viscoelastic.  Besides, VDLO can consider not only limit loads but also limit displacements.
	\item
    Body forces are automatically considered in the same manner as inner stresses without any special treatment.
	\item
	Although VDLO is a node-based method, it can be implemented in other numerical tools incorporated in the framework of element/node-based methods.
	\item
	VDLO can be employed more easily than classic DLO for pseudostatic analysis.
\end{itemize}
The remainder of this paper is organized as follows: In Section~\ref{sec:vdlo}, VDLO is presented in detail, including the balance relationship, the target function and the constraints.  For clarity symbols similar to the global cracking element method \cite{Yiming:20} are used.  In Section~\ref{sec:NEs}, results from some numerical examples referring to the Prandtl test, the uniaxial compression test of porous media, and to the Kalthoff test are provided to demonstrate the effectiveness of VDLO.  Concluding remarks are made in Section~\ref{sec:conc}.

\section{Virtual displacement-based discontinuity layout optimization}
\label{sec:vdlo}
\subsection{Balance relationship and target function}
The symbols used in this work are consistent with those used in former work of the authors referring to cracking elements \cite{Yiming:20,Yiming:23} and DLO \cite{Yiming:12,Yiming:17}.  The symbols with subscripts represent local parameters, while the symbols without subscripts represent global parameters.

\begin{figure}[htbp]
	\centering
	\includegraphics[width=0.7\textwidth]{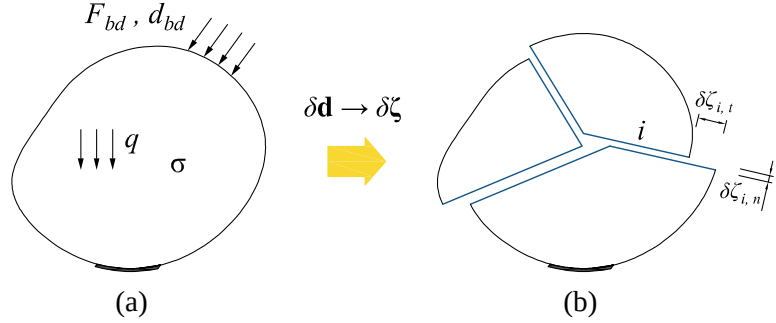}
	\caption{Domain $\Omega$: (a) stress field $\boldsymbol{\sigma}$ induced by boundary forces, boundary displacements, and body forces; (b) virtual displacement jumps (virtual crack openings)}
	\label{fig:domain}
\end{figure}

At a specific moment, a domain $\Omega$ is subjected to body forces $q$.  Its boundary $\partial \Omega$ is subjected to loads $F_{bd}$ and displacements $d_{bd}$.  This results in a stress field $\boldsymbol{\sigma}$ in $\Omega$ (see Figure~\ref{fig:domain}(a)).  Taking the stress state at the moment and assigning a virtual displacement field $\delta \mathbf{d}$ to $\Omega$ as a snapshot, the relationship $\delta W=\delta E$ holds, where $\delta W$ and $\delta E$ are the virtual driving and resistance works done.  Furthermore, considering the upper bound limit state of a perfectly plastic material, when the limit state is reached, the virtual displacements are composed only of virtual displacement jumps (virtual crack openings) $\delta\boldsymbol{\zeta}$; for the discontinuity $i$, the normal and shear forces 
are denoted $ \delta\boldsymbol{\zeta}_i=\left[\delta\zeta_{i,t},\delta\zeta_{i,n}\right]^T$ (see Figure~\ref{fig:domain}(b)).  Correspondingly, $\delta W$ and $\delta E$ are induced by $\delta\boldsymbol{\zeta}$.  Herein, $\delta W$ denotes the virtual work done by $\delta\boldsymbol{\zeta}$ and $\boldsymbol{\sigma}$ whereas $\delta E$ stands for the virtual work done by $\delta\boldsymbol{\zeta}$ and friction.

\begin{figure}[htbp]
	\centering
	\includegraphics[width=0.9\textwidth]{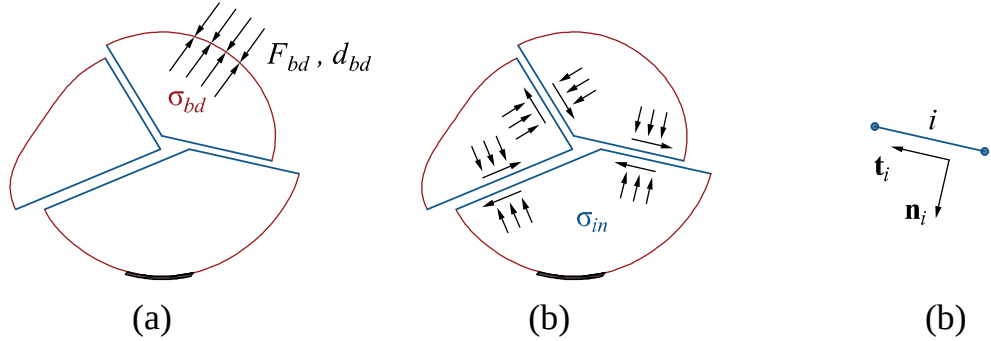}
	\caption{Domain $\Omega$ at the limit state: (a) balance of boundary forces and stresses at the boundary, (b) virtual work done by the inner stresses, and (c) unit vectors $\mathbf{t}_i$ and $\mathbf{n}_i$ of the discontinuity $i$}
	\label{fig:work}
\end{figure}

Since the boundary forces $F_{bd}$ and the boundary stresses $\boldsymbol{\sigma}_{bd}$ are equal, the virtual work done by these two quantities compensate each other (see Figure~\ref{fig:work}(a)).  Moreover, $\delta\boldsymbol{\zeta}$ at the boundary denotes displacements, rather than displacement jumps.  These $\delta\boldsymbol{\zeta}$ do not dedicate to $\delta E$.  Hence, it is only necessary to consider the inner stresses $\boldsymbol{\sigma}_{in}$ and the inner $\delta\boldsymbol{\zeta}$ when calculating $\delta W$ and $\delta E$ (see Figure~\ref{fig:work}(b)).  For convenience, in the remaining part of the manuscript $\boldsymbol{\sigma}$ represents $\boldsymbol{\sigma}_{in}$ ($\boldsymbol{\sigma}_{bd}$ will not be taken into account).  Considering the definitions of $\delta W$ and $\delta E$, the following relations are obtained:
\begin{equation}
\begin{aligned}
&\delta W=\delta E,\\
&\mbox{where}\\
&\delta W=\sum_i\left[\delta\zeta_{i,t}\left(\mathbf{n}_i\otimes\mathbf{t}_i\right):\mathbf{S}_i+\delta\zeta_{i,n}\left(\mathbf{n}_i\otimes\mathbf{n}_i\right):\mathbf{S}_i\right],\\
&\mbox{with }\mathbf{S}_i=\int \boldsymbol{\sigma}_i\ d\ l_i,\\
&\mbox{and}\\
&\delta E=\sum_i \left( | \delta\zeta_{i,t} |\ c_i\ l_i\right),
\label{eq:balance}
\end{aligned}
\end{equation}
where $\mathbf{t}_i$ and $\mathbf{n}_i$ are unit vectors parallel and perpendicular to the discontinuity $i$ \cite{Yiming:11}, with $\mathbf{t}_i=\left[t_{i,x}, t_{i,y}\right]^T$ and $\mathbf{n}_i=\left[n_{i,x}, n_{i,y}\right]^T$ (see Figure~\ref{fig:work}(c)); $c_i$ and $l_i$ denote the cohesion and the length of $i$, respectively; and $\int \cdot \ d\ \ l_i$ is the integral of the corresponding function along the discontinuity $i$.

Eq.~\ref{eq:balance} is satisfied when $\Omega$ reaches the limit state.  If the present structure does not reach this state, its factor of safety $\lambda$ can be defined as
\begin{equation}
\lambda=\frac{\delta E}{\delta W},
\label{eq:fos}
\end{equation}
where $\delta E$ and $\delta W$ may be considered as the ``resisting factor'' and ``driving factor'', respectively.  $\lambda>1$ indicates that the resisting factor is greater than the driving factor and $\Omega$ is safe or stable.  When the present $\boldsymbol{\sigma}$ (snapshot $\boldsymbol{\sigma}$) is induced by the prescribed load ``$\mathbf{F}$'' or the displacement ``$\mathbf{d}$'', then ``$\lambda\ \mathbf{F}$'' or ``$\lambda\ \mathbf{d}$'' is the limit load/displacement of the present system.  In addition, in this formulation, $\boldsymbol{\sigma}$ is assumed to be caused only by live loads.  However, a similar formulation can also be deduced for dead loads.

To obtain the optimum layout of the discontinuities, loads/displacements are set to result in $\delta W=1$ for triggering failure.  Thus, the target function becomes
\begin{equation}
\lambda=\delta E\rightarrow \mbox{min}.
\label{eq:targetfunction}
\end{equation}

It is emphasized that these deductions are based on two assumptions: i) all activated discontinuities emerge at the same time (cracks do not propagate), and ii) the failure pattern of the material does not involve strain-softening, because in such a case the activation of discontinuities would result in a considerable redistribution of stress, making the snapshot of stress irrelevant.  On the other hand, the classic DLO algorithm also follows these two assumptions.

\subsection{Constraints}
The constraints in the VDLO optimization problem are almost the same as those in the classic DLO, which will be introduced briefly in this section; more details can be found in \cite{Smith:01,Yiming:12}.
\subsubsection{Compatibility constraints}
~\\
Compatibility constraints are introduced to prevent subdomains, separated by potential discontinuities from intruding into and overlapping one another.  Considering a node $P$, connecting several discontinuities $i\cdots m$, the compatibility constraint at this node is obtained as
\begin{equation}
\mathbf{X}_P=\left[\begin{array}{ccc}
\mathbf{B}_{P,i}&\cdots&\mathbf{B}_{P,m}\end{array}\right]\left[\begin{array}{c}
\delta\boldsymbol{\zeta}_i\\
\vdots\\
\delta\boldsymbol{\zeta}_m\end{array}\right]=\mathbf{0},
\label{eq:bp}
\end{equation}
where
\begin{equation}
\mathbf{B}_{P,i}
=
\left[\begin{array}{cc}
t_{i,x} &-t_{i,y}\\
t_{i,y} &t_{i,x}\end{array}\right],
\label{eq:localB}
\end{equation}
with $\mathbf{t}$ pointing from $P$ to the opposite node at the discontinuity $i$.

Such compatibility constraints are considered for all nodes $P\cdots W$ in the domain, yielding the following global compatibility constraint equations:
\begin{equation}
\left[\begin{array}{c}
\mathbf{X}_P\\
\vdots\\
\mathbf{X}_W\\
\end{array}\right]
=\mathbf{B}\ \delta\boldsymbol{\zeta}=\mathbf{0}.
\label{eq:gbp}
\end{equation}
\subsubsection{Flow rule constraint regarding Mohr--Coulomb failure}
~\\
For all inner discontinuities (discontinuities inside the body), the associated flow rule regarding Mohr--Coulomb failure is characterized by the following relation for the inner discontinuity $i$:
\begin{equation}
\delta \zeta_{i,n}-|\delta \zeta_{i,t}|\ \mbox{tan}\ \phi_i=0,
\label{eq:mcfailure}
\end{equation}
where $\phi_i$ is the angle of friction (for dilation) of $i$.

As proposed in \cite{Smith:01}, we also introduce the plastic multiplier $\delta\mathbf{p}_i=\left[\delta p_{i+},\delta p_{i-}\right]^T$ as $\delta\zeta_{i,t}=\delta p_{i+}-\delta p_{i-}$ into Eq.~\ref{eq:mcfailure} to remove $|\cdot|$:
\begin{equation}
\left\{
\begin{array}{l}
\delta\zeta_{i,t}=\delta p_{i+}-\delta p_{i-}\\
\delta\zeta_{i,n}=\left( \delta p_{i+}+\delta p_{i-} \right) \ \mbox{tan}\ \phi_i \\
\mbox{with  }\delta p_{i+}\geq 0, \ \delta p_{i-}\geq 0.\\
\end{array}\right.,
\label{eq:plasticmultiplier}
\end{equation}
where $\delta p_{i+}$ and $\delta p_{i-}$ represent the relative clockwise and counterclockwise sliding, respectively, of the two subdomains separated by the discontinuity $i$.  $\left(\delta p_{i+}\ \delta p_{i-}\right)=0$ is automatically satisfied.

Next, Eq.~\ref{eq:plasticmultiplier} can be written in matrix form as
\begin{equation}
\begin{array}{l}
\mathbf{N}_i \left[ \begin{array}{c}
\delta\boldsymbol{\zeta}_i\\
\delta\mathbf{p}_i
\end{array}\right]=\left[ \begin{array}{cccc}
1&0&-1&1\\
0&-1&\mbox{tan}\ \phi_i&\mbox{tan}\ \phi_i
\end{array}\right] \left[\begin{array}{c}
\delta\zeta_{i,t}\\
\delta\zeta_{i,n}\\
\delta p_{i+}\\
\delta p_{i-}
\end{array}\right]=\mathbf{0}\\
\delta p_{i+}\geq 0, \ \delta p_{i-}\geq 0\end{array},\\
\label{eq:mcfailureNEW}
\end{equation}
as a typical constraint for linear programming (LP) problems \cite{Dantzig:01}.

For all inner discontinuities in the system, a global flow rule constraint is obtained:
\begin{equation}
\mathbf{N} 
\left[ \begin{array}{c}
\delta\boldsymbol{\zeta}\\
\delta\mathbf{p}
\end{array}\right]=\mathbf{0}.
\label{eq:mcfailureAll}
\end{equation}
It is again emphasized that only the inner discontinuities are considered in $\mathbf{N}$.

\subsubsection{Unit virtual work constraint}
~\\
As mentioned previously, $\delta W=1$ represents the driving factor.  When using Voigt's notation for representing symmetric second-order tensors and considering the stress along an inner discontinuity $i$ as $\boldsymbol{\sigma}_i=\left[\sigma_{i,x},\sigma_{i,y},\tau_{i,xy}\right]^T$.  The unit virtual work constraint can be written for all inner discontinuities $i\cdots j$ as
\begin{equation}
\begin{aligned}
&\delta W=\mathbf{G}\ \delta\boldsymbol{\zeta}=\left[\begin{array}{ccc}G_{i,t},G_{i,n}\cdots G_{j,t},G_{j,n}\end{array}\right]\left[\begin{array}{c}
\delta\zeta_{i,t}\\
\delta\zeta_{i,n}\\
\vdots\\
\delta\zeta_{j,t}\\
\delta\zeta_{j,n}\end{array}\right]=1,\\
&\mbox{where}\\
&G_{i,t}=\int \left[\sigma_{i,x},\sigma_{i,y},\tau_{i,xy}\right]\left[\begin{array}{c}
t_{i,x}\ n_{i,x}\\
t_{i,y}\ n_{i,y}\\
t_{i,x}\ n_{i,y}+t_{i,y}\ n_{i,x}\end{array}\right]\ d\ l_i,\\
&G_{i,n}=\int \left[\sigma_{i,x},\sigma_{i,y},\tau_{i,xy}\right]\left[\begin{array}{c}
n^2_{i,x}\\
n^2_{i,y}\\
2\ n_{i,x}\ n_{i,y}\end{array}\right]\ d\ l_i.\\
\label{eq:deltaWeq1}
\end{aligned}
\end{equation}

\subsection{Optimization equation}
Considering Eqs.~\ref{eq:gbp},~\ref{eq:mcfailureAll}, and~\ref{eq:deltaWeq1}, VDLO optimization (in matrix form) is formulated as follows:
\begin{equation}
\begin{array}{l}
\mbox{minimize }\ \ 
\delta E=\left[c_i\ l_i\cdots c_j\ l_j\right]\left[\begin{array}{c}
\delta\mathbf{p}_i\\
\vdots\\
\delta\mathbf{p}_j\end{array}\right]\\
\\
\mbox{subject to }\\
\\
\left[\begin{array}{ccc}
\mathbf{B}&&\mathbf{0}\\
&\mathbf{N}&\\
\mathbf{G}&&\mathbf{0}
\end{array}\right]\left[\begin{array}{c}
\delta\boldsymbol{\zeta}\\
\delta\mathbf{p}
\end{array}\right]=\left[\begin{array}{c}
\mathbf{0}\\
\mathbf{0}\\
1
\end{array}\right]\\
\\
\mbox{and }\\
\\
\left\{
\begin{array}{l}
\forall i \in I_t, \delta\zeta_{i,t}=0\\
\forall i \in I_n, \delta\zeta_{i,n}=0\\
\delta\mathbf{p}\geq \mathbf{0}\\
\end{array}\right. ,
\end{array}
\label{eq:DLOMa}
\end{equation}
where $I_t$ and $I_n$ are the sets of indices of discontinuities fixed along the shear and normal directions, respectively.

\subsection{Implementation in the FEM framework}
VDLO requires a stress field in the domain.  It can be provided by any numerical tool. Because the FEM is the most commonly used numerical tool for computer-aided design (CAD) and computer-aided engineering (CAE).  In this work, VDLO is implemented in the FEM framework.

\begin{figure}[htbp]
	\centering
	\includegraphics[width=0.8\textwidth]{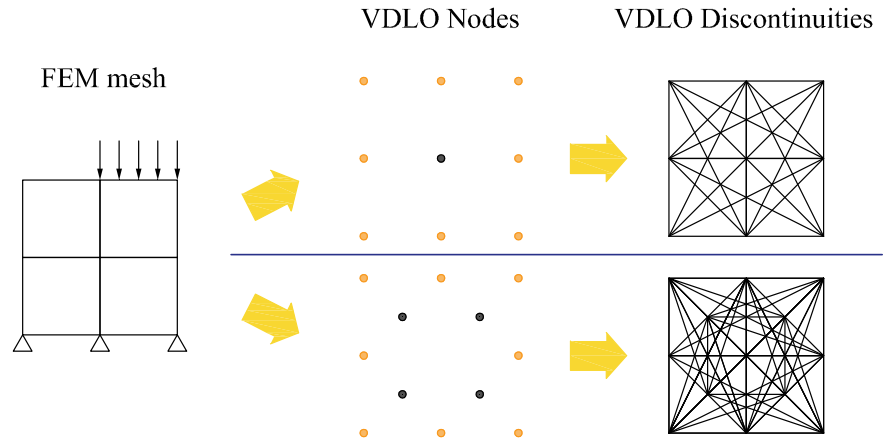}
	\caption{Two strategies for transforming the FEM model into the VDLO model: i) VDLO shares the same nodes with the FEM; ii) VDLO uses Gaussian points (only the central Gaussian points are shown in this figure for simplicity) and FEM boundary nodes}
	\label{fig:FEMtoDLO}
\end{figure}

However, the stress field is not continuous at the nodes of the FEM mesh.  There are two options for transforming the FEM model into the VDLO model:
\begin{itemize}
	\item
	First, the stresses are smoothed \cite{Hinton:01} to guarantee the continuity of the stress field at the FEM nodes; then, the same nodes are used for the VDLO model, and each pair of nodes is treated as a potential discontinuity.
	\item
	Gaussian points are used for the elements and FEM nodes are used on the boundary as VDLO nodes.  Then, each pair of nodes is treated as a potential discontinuity.
\end{itemize}
These two strategies are illustrated in Figure~\ref{fig:FEMtoDLO}.  The first strategy has two advantages: i) the same nodes are shared by the FEM and VDLO method, eliminating the need for extra storage of the VDLO nodes, and ii) Newton--Cotes integration can be used for Eq.~\ref{eq:deltaWeq1}. The main drawback, however, is that extra smoothing is needed for the stress field, which is not required by the second strategy.  In other words, although the boundary nodes of the FEM model are introduced into the VDLO model, the stresses at the boundary nodes can be ignored, because Eq.~\ref{eq:deltaWeq1} does not take boundary discontinuities into account.  Nevertheless, considering Eq.~\ref{eq:deltaWeq1}, the second strategy requires Gaussian integration for the inner discontinuities connected to any boundary node, whereas Newton--Cotes integration should be used for inner discontinuities connecting two Gaussian points.  Both strategies support adaptive refinements \cite{Yiming:17}.

In order to save storage and to use the same integration method for all discontinuities, considering Eq.~\ref{eq:deltaWeq1}, the first strategy (a global smoothing strategy) is adopted in this work.  Assembling the stresses at nodes to a vector denoted $\boldsymbol{\sigma_N}$ and the stresses at the Gaussian points to a vector denoted $\boldsymbol{\sigma_G}$, the following equation is obtained:
\begin{equation}
	\mathbf{A}\ \boldsymbol{\sigma_N}=\boldsymbol{\sigma_G},
	\label{eq:gs}
\end{equation}
where $\mathbf{A}$ is a sparse matrix with $g$ rows and $n$ columns.  $g$ is the number of Gaussian points ($g>n$ in most cases), and $n$ is the number of nodes.  $\mathbf{A}$ is determined with the help of shape functions.  Eq.~\ref{eq:gs} could be directly solved by QR decomposition.  Instead, in this work, the least squares method is used to solve the linear system $\left(\mathbf{A}^T\mathbf{A}\right)\ \boldsymbol{\sigma_N}=\mathbf{A}^T\boldsymbol{\sigma_G}$ with the Multifrontal Massively Parallel Sparse Direct Solver (MUMPS) \cite{mumps,AMESTOY2000501,AMESTOY2003833}.  Similar to the global stiffness matrix, $\left(\mathbf{A}^T\mathbf{A}\right)$ can be easily and efficiently assembled element by element.  After smoothing the stress field, Newton--Cotes integration is used for numerically calculating the integrals in Eq.~\ref{eq:deltaWeq1}.  For long discontinuities, several temporary points between the nodes are used to enhance the accuracy.

\section{Numerical examples}
\label{sec:NEs}

\subsection{Bearing capacity test}
\begin{figure}[htbp]
	\centering
	\includegraphics[width=0.95\textwidth]{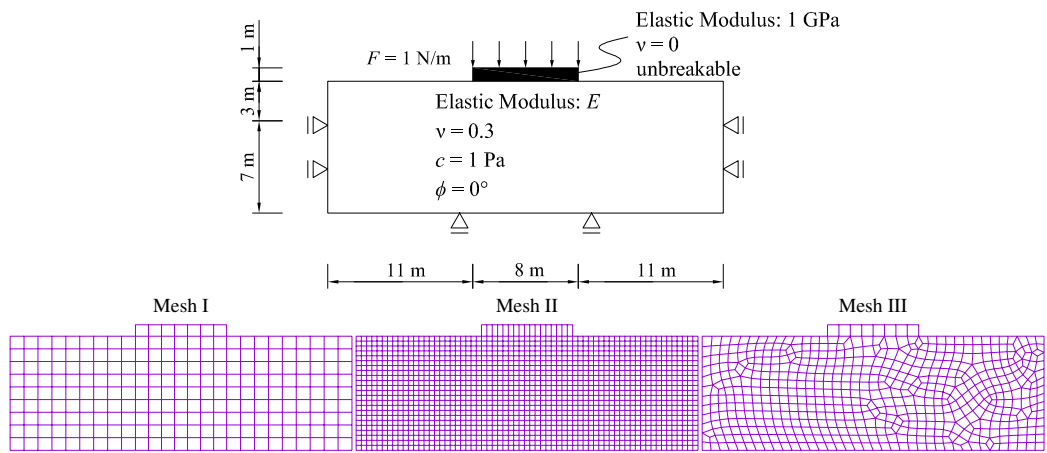}
	\caption{Bearing capacity test: model, material and mesh}
	\label{fig:Ex1Model}
\end{figure}

The first example is a classic bearing capacity test.  The plane stress condition is considered.  As shown in Figure~\ref{fig:Ex1Model}, the model allows consideration of both regular and irregular meshes.  The elastic modulus of the soil is denoted as $E$.

\begin{figure}[htbp]
	\centering
	\includegraphics[width=0.9\textwidth]{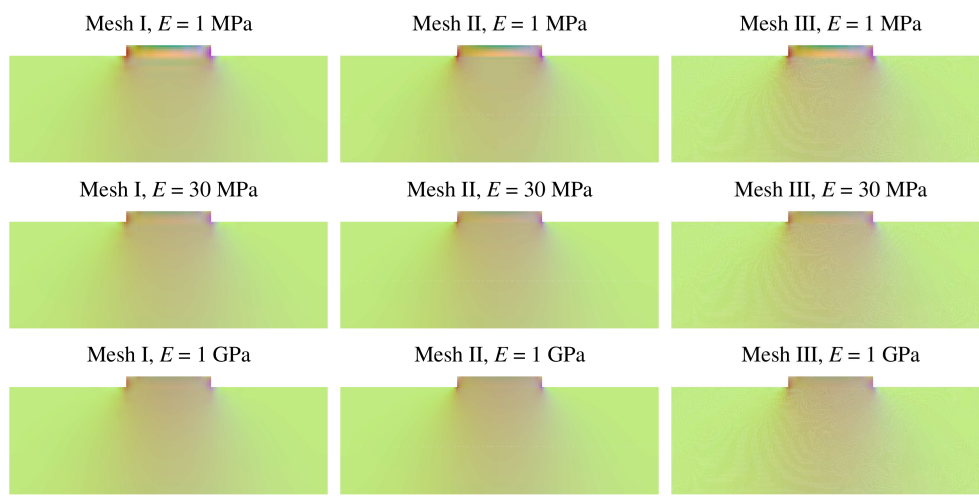}
	\caption{Bearing capacity test: smoothed stress fields $\boldsymbol{\sigma}=\left[\sigma_{x},\sigma_{y},\tau_{xy}\right]$ shown by RGB images: red for $\sigma_{x}\in\left[-15\mbox{ Pa}, 5\mbox{ Pa}\right]\rightarrow \left[0, 255\right]$, green for $\sigma_{y}\in\left[-3.5\mbox{ Pa}, 0.3\mbox{ Pa}\right]\rightarrow \left[0, 255\right]$, and blue for $\tau_{xy}\in\left[-1.5\mbox{ Pa}, 1.5\mbox{ Pa}\right]\rightarrow \left[0, 255\right]$}
	\label{fig:Ex1StressSym}
\end{figure}

Considering different values of $E$, the smoothed stress fields are illustrated by RGB images in Figure~\ref{fig:Ex1StressSym} \cite{Yiming:24}, indicating very similar smoothed stress distributions for different meshes.

\begin{figure}[htbp]
	\centering
	\includegraphics[width=0.95\textwidth]{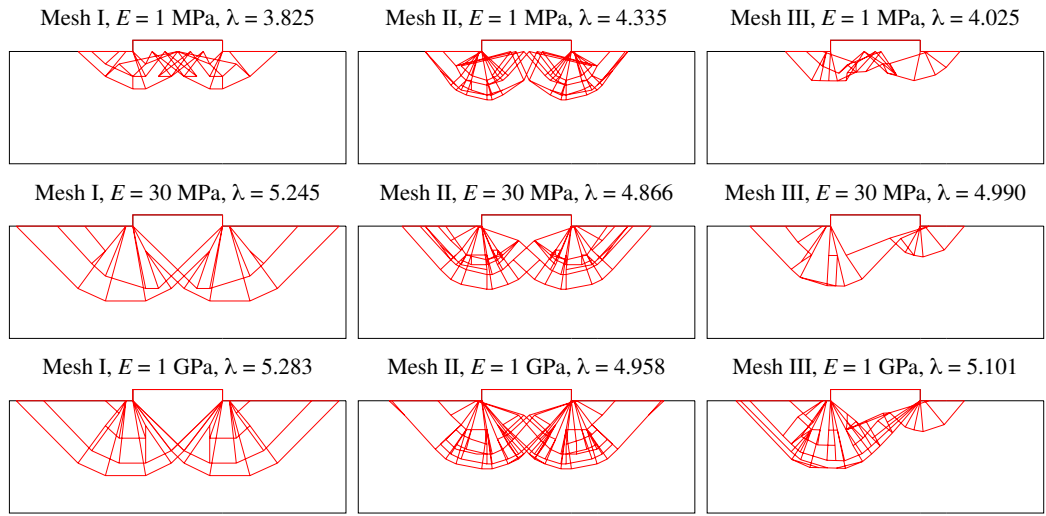}
	\caption{Bearing capacity test: factors of safety and failure patterns, considering different values of $E$ and different meshes}
	\label{fig:Ex1failure}
\end{figure}

The obtained factors of safety and failure patterns are plotted in Figure~\ref{fig:Ex1failure}.  When increasing the elastic modulus, the bearing capacity of the foundation is generally increased while the cases with $E=$30~MPa and $E=$1~GPa do not provide very different results.  The symmetric mesh provides nearly symmetric failure patterns, but some small differences can be found, mainly as a result of decimal errors when calculating Eq.~\ref{eq:deltaWeq1}.  In contrast, mesh II does not provide symmetric results because of the asymmetry of the potential discontinuities.  The obtained values of $\lambda$ are close to the analytical result of a classic Prandtl test as $2+\pi$ \cite{Prandtl:01}, and the failure patterns are acceptable.

\subsection{Strength of matrix inclusion materials}
\begin{figure}[htbp]
	\centering
	\includegraphics[width=0.9\textwidth]{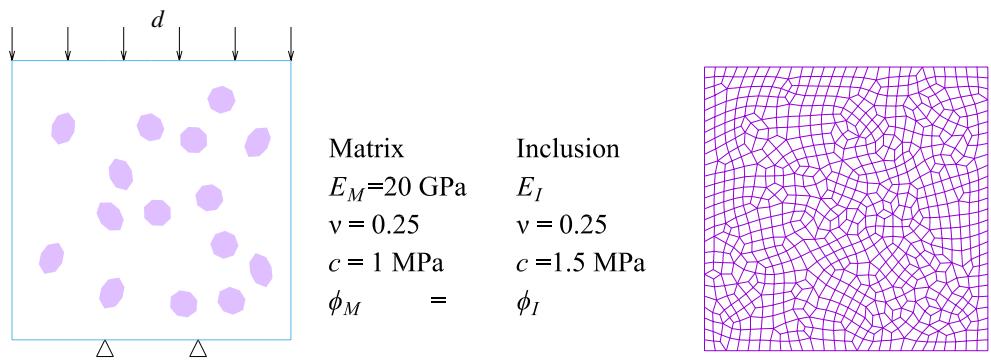}
	\caption{Matrix inclusion material: model, material and mesh}
	\label{fig:Ex2Model}
\end{figure}

DLO was used to predict the yield strength and failure patterns of randomly generated matrix inclusion materials (see, for example, \cite{BAUER201582}).  In this section, the VDLO is used to investigate the strength of the matrix inclusion material shown in Figure~\ref{fig:Ex2Model}.  Again, plane stress is considered.  The width and height of the structure are both 1~m, and the volume of every inclusion is 0.008 m$^3$.  The elastic modulus and the strength of the matrix remain fixed, whereas the properties of the inclusions are changed to check their responses.  Displacement boundary conditions are specified on the top and bottom of the model with $d=0.1\mbox{ mm}$.

\begin{figure}[htbp]
	\centering
	\includegraphics[width=1\textwidth]{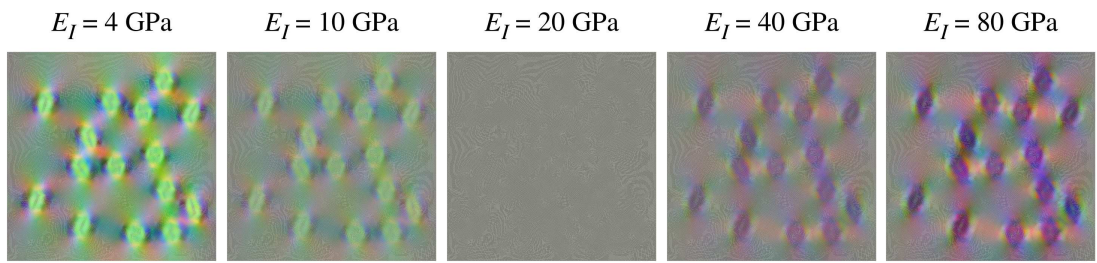}
	\caption{Matrix-inclusion material: smoothed stress fields $\boldsymbol{\sigma}=\left[\sigma_{x},\sigma_{y},\tau_{xy}\right]$ illustrated by RGB figures: red for $\sigma_{x}\in\left[-0.6\mbox{ MPa}, 0.6\mbox{ MPa}\right]\rightarrow \left[0, 255\right]$, green for $\sigma_{y}\in\left[-4\mbox{ MPa}, -0.1\mbox{ MPa}\right]\rightarrow \left[0, 255\right]$, and blue for $\tau_{xy}\in\left[-0.6\mbox{ MPa}, 0.65\mbox{ MPa}\right]\rightarrow \left[0, 255\right]$}
	\label{fig:Ex2Stress}
\end{figure}

\begin{figure}[htbp]
	\centering
	\includegraphics[width=1\textwidth]{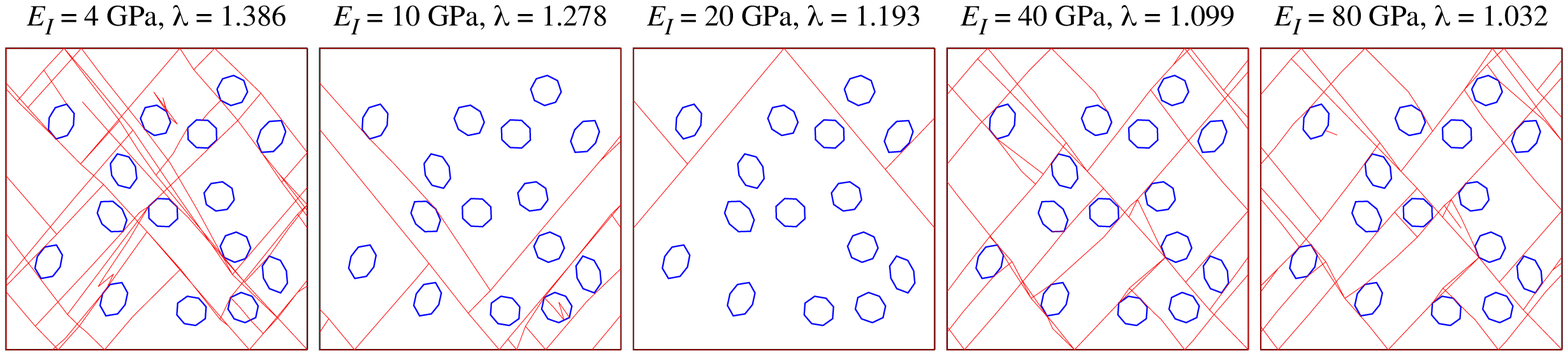}
	\caption{Matrix inclusion material: factors of safety and failure patterns for $\phi_M=\phi_I=10^\circ$}
	\label{fig:Ex2failure}
\end{figure}

\begin{figure}[htbp]
	\centering
	\includegraphics[width=0.98\textwidth]{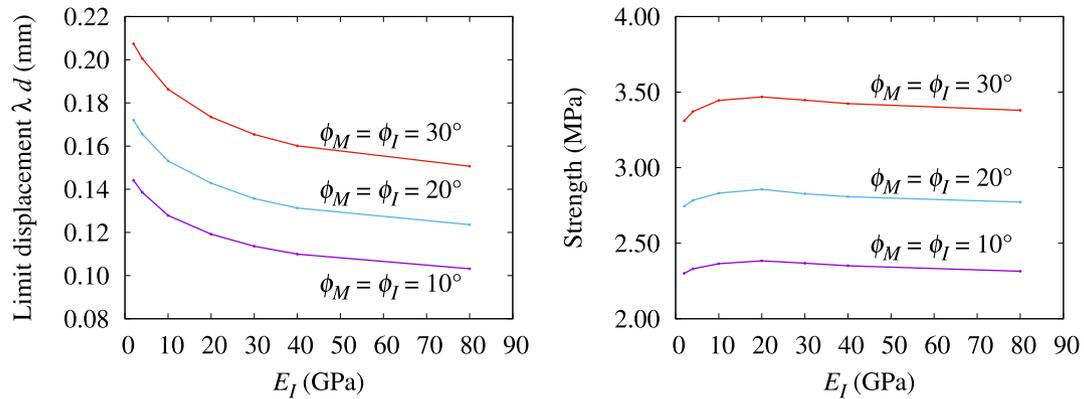}
	\caption{Matrix-inclusion material: the relationships between the limit displacement $\lambda d$ and strength}
	\label{fig:Ex2Strength}
\end{figure}

Considering different values of $E_I$, the smoothed stress fields are shown in Figure~\ref{fig:Ex2Stress}.  If $E_I\ne E_M$, the stress is obviously concentrated near the inclusions.  In addition, as $E_I$ increases, the RGB stress figures change from green to red, indicating an increase in $\sigma_x$.  When $\phi_M=\phi_I=10^\circ$, the corresponding factors of safety $\lambda$ and failure patterns are provided in Figure~\ref{fig:Ex2failure}.  Because the cohesion of the inclusions is larger than that of the matrix, most discontinuities bypass the inclusions.  Interestingly, it is noticed that more discontinuities appear if $E_I\ne E_M$.  Considering different friction angles, the relationships between the limit displacement $\lambda d$ and strength are illustrated in Figure~\ref{fig:Ex2Strength}.  The strength is greatly influenced by the friction angle while the influence of $E_M$ on the strength is not evident.  Instead, the influence of $E_M$ on the limit displacement $\lambda d$ is more pronounced.

\subsection{Kalthoff test (mode II)}
\begin{figure}[htbp]
	\centering
	\includegraphics[width=0.9\textwidth]{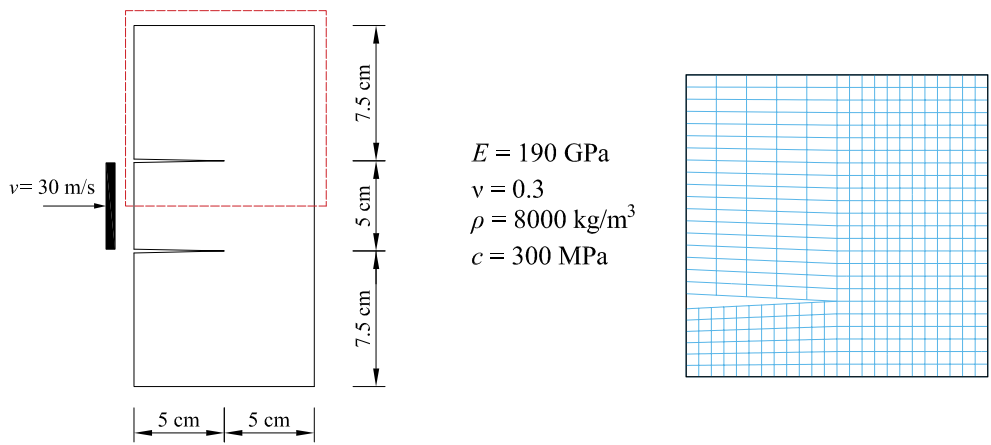}
	\caption{Kalthoff test (mode II): model, material, and mesh}
	\label{fig:Ex3Model}
\end{figure}

Next the Kalthoff test is investigated \cite{Kalthoff:01}.  The model, the material and the mesh are shown in Figure~\ref{fig:Ex3Model}.  Because of symmetry, only one half of the structure is considered.  In \cite{Kalthoff:01}, it is shown that for a sharp crack, with the impact velocity exceeding 30~m/s, the damage behavior changes from mode I to mode II.  Simulation of mode-I failure can be found in many studies (e.g., \cite{SongandBelytschko,Rabczuk2007shear,MANDAL2020107169}).  Thus, the focus in this work is on mode-II damage.  As a 2D example, the plane strain condition is assumed.  In the FEM analysis step, the Newmark method is used with the Newmark parameters $\alpha=0.25$ and $\delta=0.5$ \cite{Yiming:16}.  The time increment for the calculation is $\Delta t=0.5\ \mu\mbox{s}$.

As mentioned before, DLO-type methods permit determination of the upper bound limit/failure pattern of a structure but do not allow to consider the crack propagation process.  Therefore, this method is generally unsuitable for transient analysis.  However, VDLO takes snapshots of the stress state for further analysis and estimates the corresponding factor of safety and failure pattern of the structure at that moment, thereby providing a pseudostatic analysis \cite{Smith:03}.  In this case, the use of VDLO avoids time-consuming nonlinear continuous--discontinuous analyses while still providing inspiring results.

\begin{figure}[htbp]
	\centering
	\includegraphics[width=0.7\textwidth]{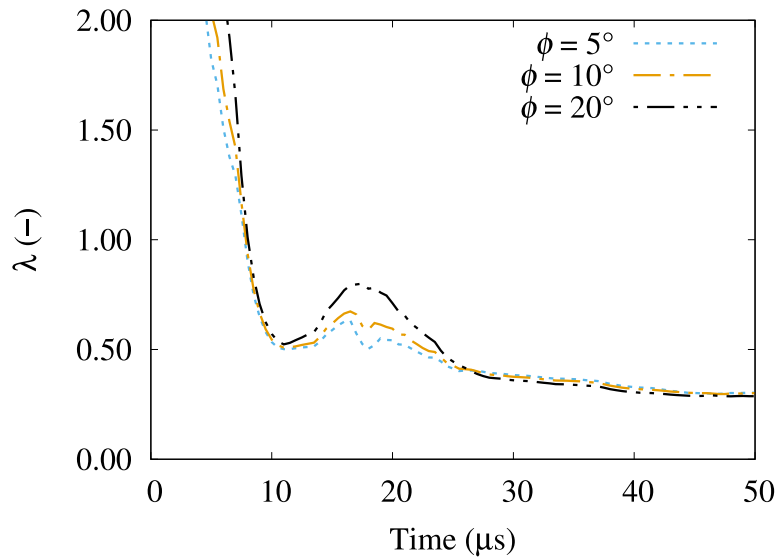}
	\caption{Kalthoff test (mode II): evolutions of $\lambda$ with time}
	\label{fig:Ex3lambda}
\end{figure}

\begin{figure}[htbp]
	\centering
	\includegraphics[width=0.86\textwidth]{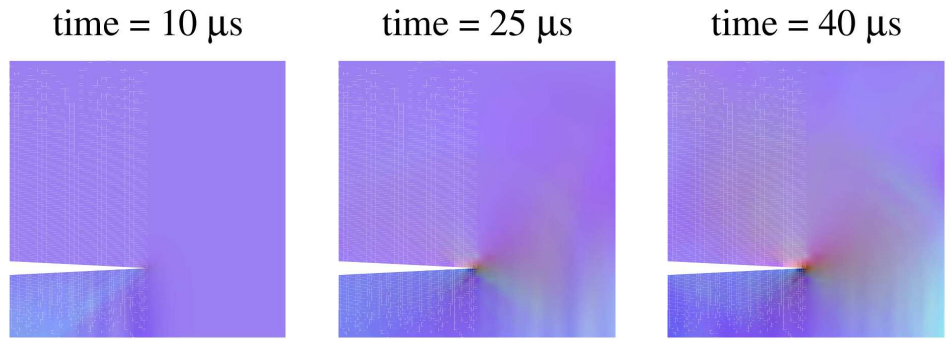}
	\caption{Kalthoff test (mode II): smoothed stress fields $\boldsymbol{\sigma}=\left[\sigma_{x},\sigma_{y},\tau_{xy}\right]$ illustrated by RGB figures: red for $\sigma_{x}\in\left[-8\mbox{ GPa}, 5\mbox{ GPa}\right]\rightarrow \left[0, 255\right]$, green for $\sigma_{y}\in\left[-2\mbox{ GPa}, 2\mbox{ GPa}\right]\rightarrow \left[0, 255\right]$, and blue for $\tau_{xy}\in\left[-3\mbox{ GPa}, 0.15\mbox{ GPa}\right]\rightarrow \left[0, 255\right]$}
	\label{fig:Ex3Stress}
\end{figure}

\begin{figure}[htbp]
	\centering
	\includegraphics[width=0.98\textwidth]{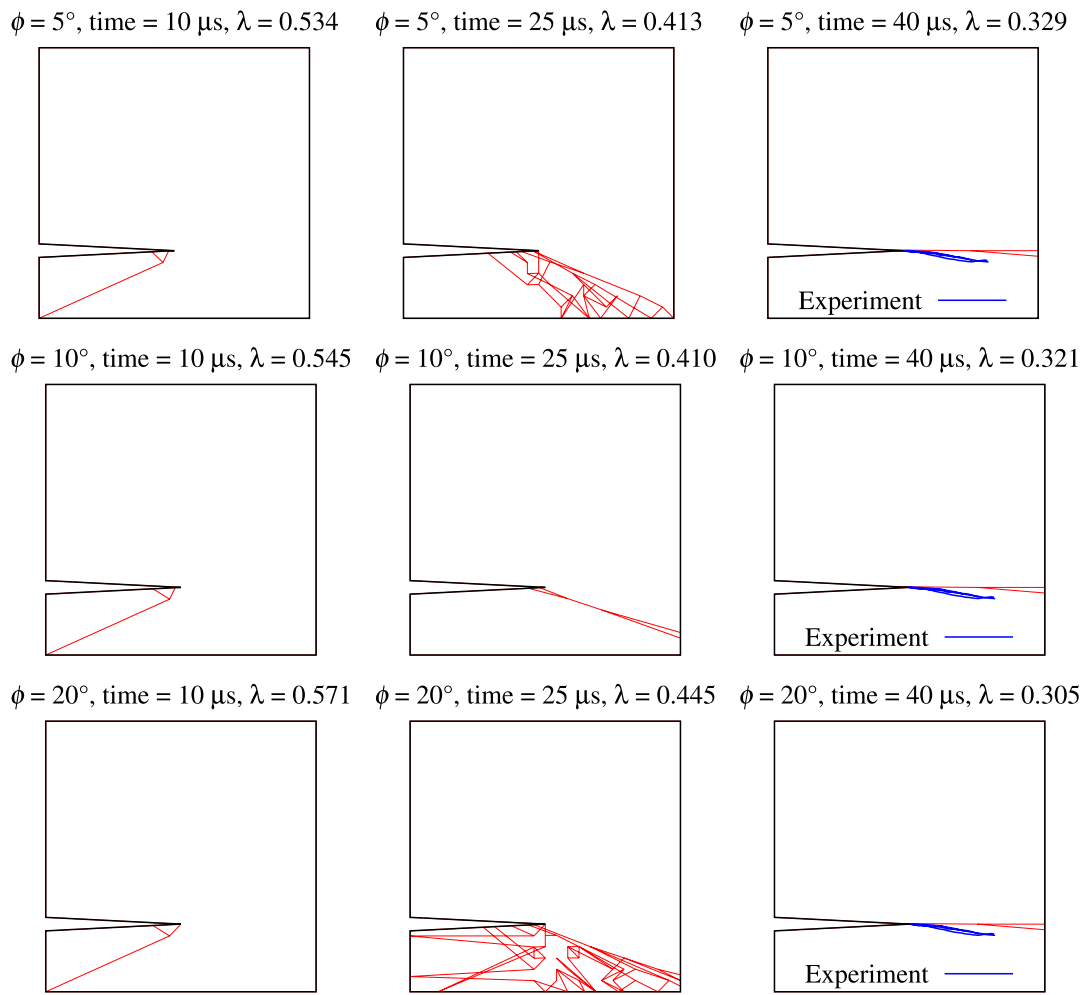}
	\caption{Kalthoff test (mode II): failure patterns and $\lambda$ at 10~$\mu\mbox{s}$, 25~$\mu\mbox{s}$, and 40~$\mu\mbox{s}$; the experimental results of \cite{Kalthoff:01} are shown for comparison}
	\label{fig:Ex3Strength}
\end{figure}

Considering different friction angles $\phi$, the evolutions of the factors of safety $\lambda$ with time are plotted in Figure~\ref{fig:Ex3lambda}.  The results indicate that $\lambda$ decreases much faster after loading.  After 10~$\mu\mbox{s}$, $\lambda$ stops decreasing sharply.  This indicates that the elastic wave reaches the crack tip at $0.5$~cm.  From 10~$\mu\mbox{s}$ to 25~$\mu\mbox{s}$, the wave propagates through the tip, causing the failure pattern to change; thus, $\lambda$ increases just slightly.  After 25~$\mu\mbox{s}$, $\lambda$ continues decreasing.  In addition, $\phi$ does not have a considerable influence on the results.  The smoothed stress fields at 10~$\mu\mbox{s}$, 25~$\mu\mbox{s}$, and 40~$\mu\mbox{s}$ are shown in Figure~\ref{fig:Ex3Stress}.  The failure patterns and $\lambda$ at these moments are shown in Figure~\ref{fig:Ex3Strength}.  As the elastic wave propagates, the main discontinuities originate at the left of the tip and propagate to the right, becoming straighter with increasing propagation.  The final shapes of the discontinuities are generally similar to the experimental results, which are almost unaffected by the friction angles.

\section{Conclusions}
\label{sec:conc}
In this work, an upper bound limit analysis method, known as VDLO, based on the classic DLO algorithm, was proposed.  VDLO uses the instantaneous stress state of the structure as its input to estimate the optimum failure pattern and factor of safety of the structure at that moment.  Hence, VDLO considers the body forces automatically.  It can be implemented as a powerful add-on to other numerical frameworks, such as the FEM.  Several numerical examples verify the effectiveness of the VDLO method, which reliably captures the failure patterns and factors of safety.

\section{ACKNOWLEDGMENT}
The authors gratefully acknowledge financial support from the National Natural Science Foundation of China (NSFC) (52178324).

%% References
%%
%% Following citation commands can be used in the body text:
%% Usage of \cite is as follows:
%%   \cite{key}         ==>>  [#]
%%   \cite[chap. 2]{key} ==>> [#, chap. 2]
%%

%% References with bibTeX database:
%\end{spacing}
%\bibliographystyle{elsarticle/model3a-num-names}

%\section*{Reference}
%\clearpage
\clearpage
\bibliographystyle{ieeetr}
\bibliography{Reference}

%% Authors are advised to submit their bibtex database files. They are
%% requested to list a bibtex style file in the manuscript if they do
%% not want to use elsarticle-num.bst.

%% References without bibTeX database:

% \begin{thebibliography}{00}

%% \bibitem must have the following form:
%%   \bibitem{key}...
%%

% \bibitem{}

% \end{thebibliography}

\end{document}